\documentclass[twocolumn,showpacs,preprintnumbers,nofootinbib,amsmath,amssymb,superscriptaddress,floatfix]{revtex4}
\usepackage{epsfig} 
\usepackage{graphicx}
\usepackage{dcolumn}
\usepackage{longtable}
\usepackage{bm}

\newcommand{\bee}{\begin{equation}}
\newcommand{\ee}{\end{equation}}
\newcommand{\beea}{\begin{eqnarray}}
\newcommand{\eea}{\end{eqnarray}}

\begin{document}


\title{Taste symmetry breaking with HYP-smeared
staggered fermions }
\author{Taegil Bae}
%
%
\affiliation{
  Frontier Physics Research Division and Center for Theoretical Physics,
  Department of Physics and Astronomy,
  Seoul National University,
  Seoul, 151-747, South Korea
}
\author{David H.~Adams}
%
%
\affiliation{
  Frontier Physics Research Division and Center for Theoretical Physics,
  Department of Physics and Astronomy,
  Seoul National University,
  Seoul, 151-747, South Korea
}
\author{Chulwoo Jung}
\email{chulwoo@bnl.gov}
\affiliation{
  Physics Department,
  Brookhaven National Laboratory,
  Upton, NY 11973, USA
}
\author{Hyung-Jin Kim}
%
%
\affiliation{
  Frontier Physics Research Division and Center for Theoretical Physics,
  Department of Physics and Astronomy,
  Seoul National University,
  Seoul, 151-747, South Korea
}
\author{Jongjeong Kim}
%
%
\affiliation{
  Frontier Physics Research Division and Center for Theoretical Physics,
  Department of Physics and Astronomy,
  Seoul National University,
  Seoul, 151-747, South Korea
}
\author{Kwangwoo Kim}
%
%
\affiliation{
  Frontier Physics Research Division and Center for Theoretical Physics,
  Department of Physics and Astronomy,
  Seoul National University,
  Seoul, 151-747, South Korea
}
\author{Weonjong Lee}
\email{wlee@phya.snu.ac.kr}
\affiliation{
  Frontier Physics Research Division and Center for Theoretical Physics,
  Department of Physics and Astronomy,
  Seoul National University,
  Seoul, 151-747, South Korea
}
\author{Stephen R. Sharpe}
\email{sharpe@phys.washington.edu}
\affiliation{
  Physics Department,
  University of Washington,
  Seattle, WA 98195-1560, USA
}
\date{\today}
\begin{abstract}
We study the impact of hypercubic (HYP) smearing on the size of
taste breaking for staggered fermions,
comparing to unimproved and to asqtad-improved staggered fermions.
As in previous studies,
we find a substantial reduction in taste-breaking compared
to unimproved staggered fermions (by a factor of 4-7 on 
lattices with spacing $a\approx 0.1\,$fm).
In addition, we observe that discretization effects of next-to-leading
order in the chiral expansion (${\cal O}(a^2 p^2)$) are markedly
reduced by HYP smearing.
Compared to asqtad valence fermions, we find that taste-breaking
in the pion spectrum is reduced by a factor of 2.5-3,
down to a level comparable to the expected size of generic
${\cal O}(a^2)$ effects.
Our results suggest that, once one reaches a lattice spacing
of $a\approx 0.09\,$fm, taste-breaking will be small enough
after HYP smearing that one can use a modified power counting
in which ${\cal O}(a^2) \ll {\cal O}(p^2)$,
simplify fitting to phenomenologically interesting quantities.
\end{abstract}
\maketitle

\section{Introduction}
\label{sec:intr}
Taste-breaking is a major concern for practical applications
of staggered fermions. For standard methods to work
(including the rooting procedure used for dynamical fermions) 
taste-breaking must vanish in the continuum limit.\footnote{%
For a recent review of
these issues see Ref.~\cite{ref:sharpe:0}.}
At non-zero lattice spacing, taste-breaking is present, and leads to
complications in interpreting and fitting the results.
It is therefore important to find variants of the staggered action
for which taste-breaking is reduced.
In this article we study one particularly simple approach,
hypercubic (HYP) smearing~\cite{ref:hyp:1},
using numerical simulations to determine the size of
non-perturbative taste-breaking, and comparing with
other types of staggered fermions.

The HYP-smeared staggered action (which we
sometimes call the ``HYP action'' for short)
is obtained by inserting
HYP-smeared links in the unimproved staggered action.
HYP smearing of links was introduced in Ref.~\cite{ref:hyp:1}
as a compact smearing approach,
and has turned out to be a practical and efficacious in
several applications.
We do not repeat the definition of HYP smearing here, except to
note that key aspects are that it involves
multiple (3 level) smearing and includes reunitarization at each step. 
For more discussion of these and
other aspects of HYP smearing, see 
Refs.~\cite{ref:orginos:1,ref:lepage:1,ref:wlee:100,ref:HISQ}.

The HYP action is not an ${\cal O}(a^2)$ improved action
in the sense of Symanzik~\cite{ref:Symanzik}: while taste-breaking
$O(a^2)$ four-fermion operators are removed at tree-level,
taste-conserving tree-level
$O(a^2)$ contributions to the quark-gluon vertices
and the quark propagator remain. 
In this respect it is inferior to the 
widely-used asqtad action~\cite{ref:orginos:1,ref:lepage:1}, 
which is tree-level ${\cal O}(a^2)$ improved.
There is considerable numerical evidence, however,
that the dominant ${\cal O}(a^2)$ effects are those due
to the taste-breaking four-fermion operators, and here
the HYP action has an advantage. The multiple smearing
in the HYP action leads to couplings between quarks and
gluons with momenta $\sim \pi/a$ that are
suppressed for a larger range of gluon momenta
and polarizations than in the asqtad action.\footnote{%
This point is very clearly explained in Ref.~\cite{ref:HISQ}.
See also the analysis in Ref.~\cite{ref:wlee:100}.}
Because of this one expects taste-breaking to continue to be
suppressed beyond tree-level with the HYP action.
Indeed, in a recent calculation, Ref.~\cite{ref:HISQ} finds that the
coefficients of taste-breaking four-fermion operators induced at
one-loop are an order of magnitude smaller for the HYP
than for the asqtad action.
In other words, the HYP action turns out to be approximately one-loop
${\cal O}(a^2)$ improved for taste-breaking effects.
This property, coupled with simplicity, are the main reasons why we
think it interesting to study the HYP action.\footnote{%
Another alternative would be to use the recently introduced n-HYP
smearing~\cite{nHYP}. Here one reunitarizes but does not project back
into SU(3). This change allows the action to be used in dynamical
simulations, at no apparent cost to the reduction in UV fluctuations.
In particular, we expect the reduction in taste-breaking to be similar
using n-HYP and our ``projected-HYP'' links.  Since, however, our
ongoing calculations are being done in a mixed-action framework (using
the freely available dynamical lattices generated with the asqtad
action), we do not need to use n-HYP links, and it is preferable to
complete our studies with the same fermion action so as to allow more
systematic comparisons at different lattice spacings.}

It should also be noted that one can have both tree-level
${\cal O}(a^2)$ improvement {\em and} approximate one-loop improvement
of taste-breaking effects. This is accomplished by
the HISQ (``highly improved staggered quark'') action~\cite{ref:HISQ}.
This action also includes a tuning of the Naik term which
removes the leading $(am)^4$ error at tree-level, thus reducing the
discretization errors for heavy quarks.
This makes it the method of choice for simulating the
charm quark with staggered fermions. For light quarks, however,
which are our interest, the lack of 
complete tree-level ${\cal O}(a^2)$ improvement may not be so important
in practice, as long as the dominant taste-breaking effects are reduced.
In particular, the remaining ${\cal O}(a^2)$ errors are of the same
parametric size as those for
non-perturbatively ${\cal O}(a)$ improved Wilson and maximally-twisted fermions,
both of which are being used for large scale simulations.
Of course, the key practical issue is the relative numerical size of
the ${\cal O}(a^2)$ errors, and we return to this point in the conclusions.

A further advantage of the HYP action is that
it avoids 
another well-known problem with unimproved staggered
fermions---the large size of the one-loop contributions to
matching factors. For example, perturbation theory for $Z_m$,
the mass renormalization factor, is very poorly convergent
with unimproved staggered fermions.
This is solved by any type of link smearing
which reduces the tree-level taste-breaking four-fermion operators,
and thus by both HYP and asqtad actions~\cite{ref:wlee:101}
(and, presumably, by the HISQ action too).
This reduction has been observed also for four-fermion operators
with the HYP action~\cite{ref:wlee:2}.

\bigskip

With these considerations as motivation, we have
undertaken a thorough study of how HYP smearing
reduces taste-breaking in non-perturbative quantities. 
The perturbative improvements described above are important,
but, if the HYP action is to be phenomenologically useful,
one must also show that non-perturbative taste-breaking is
reduced. To do this we study the splittings in the pion multiplet.
This measure of taste-breaking
 has several advantages. First, very accurate
calculations are possible, allowing small effects to be picked out.
Second, it is known from previous studies that the splittings 
between pions are larger than those between states of other spin-parities
(e.g. the vector mesons or baryons).
And, third, pion taste-breaking feeds into all quantities,
since the masses of pions of all tastes enter the
non-analytic ``chiral logarithmic'' terms 
which occur in generic chiral expansions.

It will be useful to recall the pattern of taste-splittings
within the pion multiplet.
The four tastes of staggered fermions lead to 16 tastes of flavor
non-singlet pions.  These have the spin-taste structure $(\gamma_5
\otimes \xi_F)$ with $\xi_F \in \{I, \xi_5, \xi_\mu, \xi_{\mu 5},
\xi_{\mu\nu} = \frac{1}{2} [\xi_\mu, \xi_\nu]\}$.
For zero spatial momentum (which is the case we study numerically)
these states fall into 8 irreducible representations (irreps)
of the lattice timeslice group~\cite{ref:golterman:1},
with tastes
$\{I\}$, $\{\xi_5\}$, $\{\xi_i\}$, $\{\xi_4\}$, $\{\xi_{i5}\}$,
$\{\xi_{45}\}$, $\{\xi_{ij}\}$, and $\{\xi_{i4}\}$.
That with taste $\xi_5$ is the Goldstone pion corresponding to
the axial symmetry which is exact when $m=0$, and is broken spontaneously.
The properties of the pion spectrum can be studied using
staggered chiral perturbation theory \cite{ref:wlee:1,ref:bernard:1}.
It is shown in Ref.~\cite{ref:wlee:1} that, at leading order
in a joint expansion in $p^2\sim m_q$ and $a^2$,
the pion spectrum respects an SO(4) subgroup of the full SU(4) taste symmetry.
In other words, the taste symmetry breaking happens in two steps:
at ${\cal O}(p^2) \approx {\cal O}(a^2)$ the SU(4) taste symmetry
is broken down to SO(4), while at
${\cal O}(a^2 p^2)$ the SO(4) taste symmetry is broken down to
the discrete spin-taste symmetry $SW_4$~\cite{ref:wlee:1}.
As a consequence of this analysis, we expect that, to good approximation,
the pions will lie in 5 irreps of SO(4) taste symmetry, with tastes
$\{I\}$, $\{\xi_5\}$,
$\{\xi_\mu\}$, $\{\xi_{\mu5}\}$, $\{\xi_{\mu\nu}\}$.
%

A secondary motivation for this work is that the
masses of pions of all tastes are needed in our companion
calculation of $B_K$ and related matrix elements.
In particular, in staggered chiral perturbation theory
loop contributions from non-Goldstone pions 
(those with $\xi_F \ne \xi_5$) is much larger
than that from the Goldstone pions~\cite{ref:sharpe:1}.
Thus we also use this work to test different choices
of sources and sinks.

Previous studies have shown that link smearing leads
to substantial reduction in the splittings in the
pion multiplet compared to the unimproved
staggered action~\cite{ref:hyp:1,ref:orginos:1,ref:mason:1}.
Other features of the action, and in particular whether
it is fully ${\cal O}(a^2)$ improved at tree-level, have
little effect on these splittings. For example, the splittings
are reduced similarly by the asqtad action
(which contains the Naik term) and the smeared p4 action
(which contains the ``Knight's move'' 
term~\cite{ref:p4:1})~\cite{ref:p4tests}.
The type of smearing used is, however, important.
HYP smearing (or the double-smearing used in
the HISQ action) has been found to be significantly
more effective than that used in the asqtad action.
For example, Ref.~\cite{ref:HISQ} finds that the pion
taste splittings\footnote{%
The quantitative measures used are the splittings within
the pion multiplet, $\Delta (a^2 m_\pi^2)$, in the chiral limit.
We use this measure below.}
are reduced by about a factor of two
when using the HYP (or HISQ) action compared to the asqtad
action, on quenched lattices with $a\approx 0.1\;$fm.

Our aim in this paper is to confirm and extend previous studies
of HYP smearing. We focus on two aspects.
First, a study of the impact of smearing on the
relative size of different contributions in the joint
chiral-continuum expansion. We are particularly interested
in the appropriate power-counting to use for valence HYP smeared
staggered fermions.
And, second, a detailed comparison with asqtad fermions,
using lattices with light dynamical sea quarks
(which are the lattices we are using to calculate
phenomenologically important quantities).

This article is organized as follows.
The next two sections describe some technicalities of our calculation.
First, 
in Sec.~\ref{sec:cubic}, we explain how we project onto 
representations of the timeslice group
using ``cubic sources''.
Second, in Sec.~\ref{sec:op}, we describe how we construct sink operators
using two different methods, and give examples of the fitting
of pion correlators.
We then turn to our numerical results.
Sections~\ref{sec:pion:unimp} and \ref{sec:pion:hyp} present,
respectively, the pion spectrum calculated
using unimproved and HYP-smeared fermions on quenched lattices,
by comparing which we study the impact of HYP smearing on
taste-breaking, and the nature of the chiral-continuum expansion.
Section~\ref{sec:cmp:asqtad-hyp} contains our central results: a
comparison of asqtad and HYP-smeared valence fermions
calculated on $2+1$ flavor dynamical lattices.
We close with some conclusions. 
An appendix discusses the decomposition of hypercubic sink
operators into timeslice representations.

Preliminary results from this work were presented in
Ref.~\cite{ref:wlee:200,ref:wlee:102,ref:wlee:11}.

\section{Choice of sources}
\label{sec:cubic}

In order to select a specific pion taste we must
choose sources and sinks that lie in irreps
of the time-slice group.
Here we discuss how this is done for the sources. We consider
only sources with vanishing physical three-momentum.

We adopt two different methods, which we label the ``cubic U(1)'' and
the ``cubic wall'' sources. 
Both are fairly standard~\cite{ref:GGKS,ref:KP} but, for
completeness, we describe them briefly. At the end we discuss
their relative merits.

We first fix the configurations to Coulomb gauge.
Propagators are obtained, as usual, by solving the
Dirac equation with source $h$:
\begin{eqnarray}
& & (D + m) \chi = h \\
&\Rightarrow & \chi(x,a;t;\vec{A}) =
\sum_{y,b} G(x,a;y,b) h(y,b;\vec{A})
\end{eqnarray}
where $G$ is the point-to-point quark propagator, 
$x,y$ label lattice sites,
and $a,b$ are color indices.
The cubic U(1) source at timeslice $t$ is
\begin{eqnarray}
& & h(y,b;t;\vec{A}) = \delta_{y_4,t}
\sum_{\vec{n}} \delta^3_{\vec{y}, 2\vec{n}+\vec{A}} \ \eta(\vec{n},b)
\,,
\end{eqnarray}
where $\vec{n}$ is a vector labeling $2^3$ cubes in the timeslice, 
$\vec{A} \in \{(0,0,0), (1,0,0), \cdots, (1,1,1)\}$
labels points within the cubes, and
$\eta(\vec{n},b)$ is a $U(1)$ noise vector normalized so that
\begin{eqnarray}
& &
\lim_{N \rightarrow \infty} \frac{1}{N}
\sum_{\eta} \eta(\vec{n},c)  \eta^{*}(\vec{n}',c')
= \delta_{\vec{n},\vec{n}'} \delta_{c,c'}\,,
\end{eqnarray}
with $N$ the number of random vectors.
In practice we use only a single noise vector on each
configuration.
Note that the noise vector is the same irrespective of
the position in the cube, but differs from cube to cube,
and color to color.
The cubic wall source differs by having the same
noise vector on all cubes, varying only
from color to color:
\begin{eqnarray}
& & h(y,b;t;\vec{A}) = \delta_{y_4,t}
\sum_{\vec{n}} \delta^3_{\vec{y}, 2\vec{n}+\vec{A}} \ \eta(b) \,.
\end{eqnarray}

There are eight sources of each type, depending
on the choice of $\vec A$.
By combining these appropriately,
one can project onto irreps of the timeslice group.
This works as follows (using cubic U(1) sources
and $t=0$ for illustration).
Suppose that we want to select a state with spin-taste
$\gamma_S \otimes \xi_F$.
We calculate propagators from all eight sources, and then
combine them 
%
%
%
using the spin-taste matrix~\cite{ref:klu:1}
\begin{eqnarray}
& & ( \overline{ \gamma_S \otimes \xi_F } )_{C, D}
= \frac{1}{4} {\rm Tr} (\gamma_C^{\dagger} \gamma_S 
\gamma_D \gamma_F^{\dagger} ) 
\label{eq:kludef}
\end{eqnarray}
(where, following Ref.~\cite{ref:DS},
$\gamma_{\vec C}=\gamma_{C_1} \gamma_{C_2} \gamma_{C_3}$,
etc.,)
in the following way:
\begin{eqnarray}
& & \frac{1}{N} \sum_{\eta} \sum_{\vec{A}, \vec{B}} 
\chi^{*}(x,a;0;\vec{A}) \chi(y,c;0;\vec{B})
\epsilon(x) \epsilon(\vec{A}) 
( \overline{ \gamma_S \otimes \xi_F } )_{\vec{B}, \vec{A}} 
\nonumber \\
& & = \frac{1}{N} \sum_{\eta}
\sum_{\vec{n},\vec{A},b} G^{*}(x,a; [0,2\vec{n}+\vec{A}],b) \eta^{*}(\vec{n},b)
\nonumber \\ & & \hspace*{2pc}
\sum_{\vec{m},\vec{B},d} G(y,c;[0,2\vec{m}+\vec{B}],d) \eta(\vec{m},d)
\nonumber \\ & & \hspace*{2pc}
\epsilon(x) \epsilon(\vec{A}) 
( \overline{ \gamma_S \otimes \xi_F } )_{\vec{B}, \vec{A}} 
\nonumber \\
& & = \sum_{\vec{n},\vec{A},\vec{B},b} G^{*}(x,a;[0,2\vec{n}+\vec{A}],b) 
G(y,c;[0,2\vec{n}+\vec{B}],b)
\nonumber \\ & & \hspace*{2pc}
\epsilon(x) \epsilon(\vec{A})
( \overline{ \gamma_S \otimes \xi_F } )_{\vec{B}, \vec{A}} 
\nonumber \\
& & = \sum_{\vec{n},\vec{A},\vec{B},b} G(y,c;[0,2\vec{n}+\vec{B}],b) 
( \overline{ \gamma_S \otimes \xi_F } )_{\vec{B}, \vec{A}}
\nonumber \\ & & \hspace*{2pc}
G([0,2\vec{n}+\vec{A}],b;x,a)\,.
\label{eq:src:op}
\end{eqnarray}
The factors of
$\epsilon(x) = (-1)^{\sum_\mu x_\mu}$
are needed to hermitian-conjugate the 
second propagator in eq.~(\ref{eq:src:op}).
As one can see, 
the final result is
the insertion of a bilinear operator 
at the source, with quark and antiquark fields spread out over a
$2^3$ cube, made gauge invariant by the choice of
Coulomb gauge, and projected onto zero three-momentum.
The sum over $\vec{A}$ and $\vec{B}$ picks out a bilinear
lying in a particular time-slice representation.

In fact, this is not quite correct, because the division of the
timeslice into a particular set of $2^3$ cubes breaks the
single-site translation symmetry. A similar issue arises for
the sinks, and is discussed in more detail in the following
section and in Appendix~\ref{app:anatomy}.
The upshot is that the cubic U(1) source couples most strongly to
the pion with the desired taste, but has an additional coupling to 
spin-scalar states with different tastes (and which live in a
different irreps of the timeslice group).

This issue does not arise for the cubic wall sources~\cite{ref:golterman:1,ref:GGKS}. 
Here the bilinear
that one is inserting at the source is constructed by first summing
the fields at a particular position in the cubes over the timeslice,
and then combining them into the desired spin-taste. This projects
onto a true irrep of the timeslice group.

We recall that irreps of the time-slice group contain two types
of states~\cite{ref:GS}:
those which propagate in time with an alternating sign $(-1)^t$,
and those which do not.
This is related to the fact that the bilinears onto which
we project in Eq.~(\ref{eq:src:op}) involve only a sum over
a spatial cube, rather than the full $2^4$ hypercube.
Thus the timeslice operator couples to states with spin-taste
$\gamma_S \otimes \xi_F$ and to their ``time-parity partners'' with
spin-taste
$\gamma_{45S} \otimes \xi_{45F}$,
the latter propagating with an alternating sign.
(Here we use a shorthand notation exemplified 
by $\gamma_{45S}= \gamma_4 \gamma_5 \gamma_S$.)
If we are considering the pion channel then $\gamma_S=\gamma_5$ or
$\gamma_{45}$, and the alternating state is a scalar,
created respectively by $\gamma_S=\gamma_4$ and ${\bf 1}$.
These ``extra'' states do not present a significant problem---their
contribution is reduced relative to the desired pion signal because
the scalars are heavier, and can be separated because of the different
time dependence. In addition, it can be reduced by using appropriate
sinks, as discussed below. A final reduction comes from the fact
that operators with $\gamma_S=\gamma_4$ (at zero momentum and with
degenerate quarks) do not couple to scalar states in the
continuum limit since they correspond to conserved charges.

While our single timeslice sources can produce pions of all 
tastes~\cite{ref:golterman:1}, the strength of coupling to
the pion is taste-dependent.\footnote{%
We note in passing that not all irreps can be produced
by single-timeslice sources~\cite{ref:golterman:1}, but 
all pion irreps can.}
As can be seen from
Eq.~(\ref{eq:src:op}) and the definition (\ref{eq:kludef}), single-timeslice
sources imply that $S_4=F_4$. For pions having tastes with $F_4=1$
(i.e. tastes $\xi_5$, $\xi_4$, $\xi_{i5}$ and $\xi_{i4}$), 
the corresponding spin matrix is thus $\gamma_5$, 
while for those with $F_4=0$ (tastes ${\bf
1}$, $\xi_i$, $\xi_{45}$ and $\xi_{ij}$), the spin matrix is
$\gamma_S=\gamma_{45}$.  We label the former tastes LT (for ``local in
time'') and the latter NLT (``non-local in time''---because the
corresponding bilinear with the preferred spin $\gamma_5$ 
involves quark and antiquarks on different timeslices).  In the continuum
limit, operators with spin $\gamma_{45}$ have an overlap with the
state that is suppressed by $\sim m_\pi/\Lambda$ compared to that for
spin $\gamma_5$, where $\Lambda\sim \langle \bar q q\rangle/f_\pi^2$
is a QCD scale that turns out to be large.  Thus we expect that
our sources are less efficient at producing NLT pions than the LT
pions.

The two sources we use represent two extremes: the cubic U(1) source
corresponds to using a local bilinear (local at the scale of a
hypercube) summed over the time-slice, while the cubic wall source
gives rise to a bilinear composed of quarks and antiquarks which are
separately spread out across the entire time-slice.  The relative
efficacy of the sources depends on several factors: which has a better
overlap with the desired pion, and smaller overlap with excited pions
and unwanted scalars; which is more effective at reducing the noise
due to the U(1) random vectors; and whether additional noise is
introduced by the fact that the cubic source is not a true irrep of
the timeslice group.  In the absence of any clear theoretical argument
favoring one source over the other, we study the issue numerically.

\section{Choice of sinks and fitting}
\label{sec:op}
We use two different methods to construct bilinear operators at the sink: 
the  ``hypercube'' or HPC method~\cite{ref:klu:1} and
the ``Golterman method''~\cite{ref:golterman:1}. 
In the former, the bilinear operators (at zero spatial momentum) are
\begin{equation}
O_{S,F}(t) = \sum_{\vec n} \sum_{A,B} \bar{\chi}(z\!+\!A)
\overline{(\gamma_S \otimes \xi_F)}_{A,B}  \chi(z\!+\!B) 
\,,
\label{eq:HPCop}
\end{equation}
where $\vec n$ labels $2^4$ hypercubes straddling timeslices
$t$ and $t+1$, with their origins at $z=(2\vec n, t)$,
while $A$ and $B$ are now vectors in the full hypercube.
We stress that these HPC sinks always run over two timeslices, even
when the operator is local in time.
They are made gauge invariant
by fixing each timeslice to Coulomb gauge, so
that spatial links are not required, and inserting appropriate
time-directed links if $S_4\ne F_4$.
More precisely, we average over the insertion of
the time-directed link at the position of the quark and of the antiquark.
The links that are
inserted are HYP-smeared when we are using HYP-smeared fermions.

In the Golterman method, we follow the construction of Ref.~\cite{ref:golterman:1}.
For example, if $S_4=F_4$, so that the bilinear is local in time,
it is defined as
\begin{eqnarray}
& & O_{S,F}(t) = \sum_{\vec{x}} \Gamma_{S,F}(x) \bar{\chi}(x) \Omega_{S,F} \chi(x) 
\label{eq:Goltdef}\\
& & \Omega_{S,F} \ \chi(x) = 
\nonumber \\ & & \hspace*{0.5pc}
\prod_{\mu=1,2,3}
\Big[ (1 - |S_\mu - F_\mu|) + |S_\mu - F_\mu| \Phi_\mu \Big] \chi(x) \\
& & \Phi_\mu \,\chi(x) = \frac{1}{2} \Big[ \chi(x + \hat{\mu})
+ \chi(x - \hat{\mu}) \Big]
\label{eq:Phimudef}
\end{eqnarray}
where $x = (\vec{x},t)$, and $\Gamma_{S,F}(x)$ is a phase factor given in
Ref.~\cite{ref:golterman:1}.
Gauge invariance is maintained as for the HPC operators.
Note that these are single timeslice operators, unlike the corresponding
HPC operators, which are spread over two timeslices (even though the
terms contributing do not involve time links).

If $S_4\ne F_4$, the operators necessarily involve two timeslices,
and are constructed following the prescription of Ref.~\cite{ref:golterman:1},
which we do not repeat here.
Time-directed gauge links added as for the HPC operators. 

The HPC method is, perhaps, conceptually simpler, but gives bilinears
that are only approximate representations of the timeslice group,
whereas the Golterman method has the advantage of giving true
irreps~\cite{ref:golterman:1}.  In fact, the HPC operators include the
Golterman operators as the leading term in an expansion in $a$, with
subleading contributions containing derivatives and having different
spin-tastes. This is discussed in more detail in Appendix
\ref{app:anatomy}.  One thus expects the HPC sinks to lead to noisier
results with more contamination from other states,
particularly for the cubic U(1) sources which have a similar
coupling to additional representations.
We study this issue numerically in the next section.

As noted in the previous section, the pions can be divided into those
having ``LT tastes'', which can be created by sources with the
preferred choice $\gamma_S=\gamma_5$, and those with ``NLT tastes'',
for which we must use sources with $\gamma_S=\gamma_{45}$.  At the
sink end, we always choose $\gamma_S=\gamma_5$, so that the bilinears
used for the LT pions are local in time, while those used for the NLT
pions are non-local in time.

\bigskip
We now briefly describe our fitting method and give examples of
the quality of our data.
For the pions with LT tastes, the sink and source have the same 
spin-taste, so that the two-point correlation
functions have positive time-reflection parity.
Thus we fit the data to the following form:
\begin{eqnarray}
f_+(t) &=& Z_1 [\exp(-m_1 t)\! +\! \exp(-m_1(L_t\!-\!t)) ]
\nonumber \\
&+& Z_2 (-1)^t [\exp(-m_2 t)\! +\! \exp(-m_2(L_t\!-\!t)) ] 
\label{eq:LTform}
\end{eqnarray}
where $L_t=64$ is the lattice size in the time direction.
The parameters $Z_1$ and $m_1$ describe the pion contribution
(spin-taste $\gamma_5 \otimes \xi_F$), while $Z_2$ and $m_2$ describe
their time-parity partners (with spin-taste $\gamma_{4} \otimes
\xi_{45F}$).  In this study we are concerned only with the pion
spectrum, and so do not include excited states in the fit function,
but rather begin fits at times large enough that the form
(\ref{eq:LTform}) is adequate. The typical fitting range is $10
\lesssim t \lesssim 20$, with the source at $t=0$. We use uncorrelated
fits (i.e. keeping only the diagonal components of the correlation
matrix), with errors estimated by the jackknife method.

\begin{figure}[t!]
\includegraphics[width=20pc]{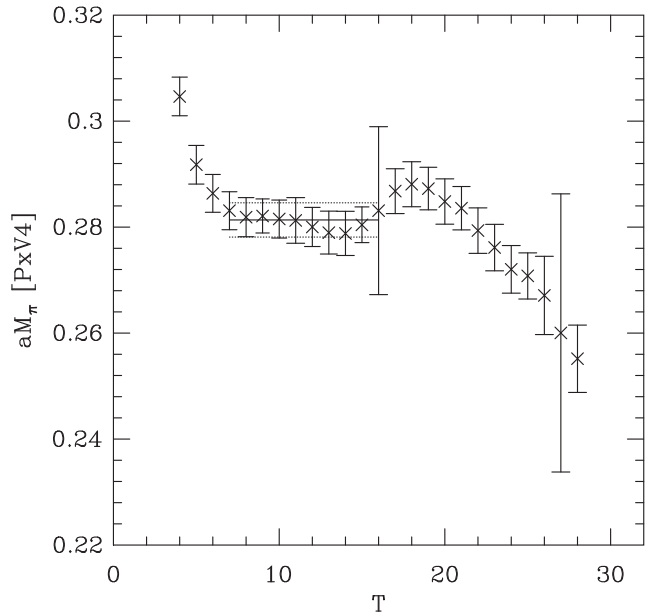}
\caption{Effective pion mass for the three-link LT pion,
$a m_\pi (\gamma_5 \otimes \xi_4)$,
for HYP-smeared staggered fermions, cubic U(1) sources (at
$t=0$), and HPC sinks. The fit is discussed in the text.
Based on 370 quenched gauge configurations
($\beta=6$), and using $m_1=m_2=0.03$ (corresponding to a particle
with mass similar to that of the physical kaon).}
\label{fig:eff-mass:s5xt4}
\end{figure}

We show an example of the data for LT pions in Fig.~\ref{fig:eff-mass:s5xt4}.
The effective mass $m_1(t)$ is obtained by equating
the form (\ref{eq:LTform}) to the lattice data for times $t-(t+3)$.
The result of a fit to the range $7 \le t \le 16$ is also shown.
A good plateau is seen, with errors of $\approx 1.4\%$.
Note that this is a ``three-link'' pion (quark and antiquark separated
by three spatial links), which typically has the 
poorest signal of the LT tastes.

\begin{figure}[t!]
\includegraphics[width=20pc]{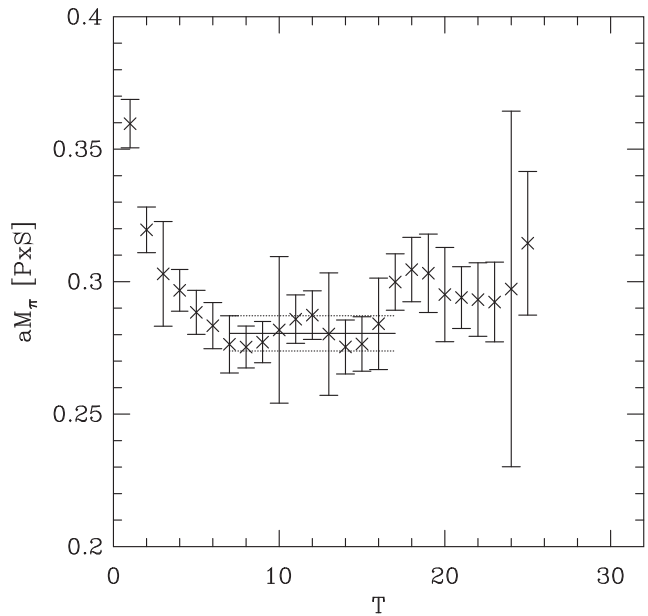}
\caption{Effective mass for four-link NLT pion,
$a m_\pi (\gamma_5 \otimes 1)$. Parameters are the same
as in Fig.~\protect\ref{fig:eff-mass:s5xt4}.}
\label{fig:eff-mass:s5xt1}
\end{figure}

For pions with NLT tastes, our sources and sinks have different spins,
and consequently the correlation functions have 
negative time-reflection parity.
The appropriate fitting function is thus
\begin{eqnarray}
f_-(t) &=& Z_1 [\exp(-m_1 t) \!-\! \exp(-m_1(L_t\!-\!t)) ]
\nonumber \\
&+& Z_2 (-1)^t [\exp(-m_2 t) \!-\! \exp(-m_2(L_t\!-\!t)) ] \,.
\nonumber 
\end{eqnarray}
An example of the corresponding effective mass is shown in
Fig.~\ref{fig:eff-mass:s5xt1}. There is a reasonable plateau, but with
errors of $\approx 3\%$ that are larger than those in 
Fig.~\ref{fig:eff-mass:s5xt4}.
 We note that this is
the pion whose bilinear is the most non-local (involving three spatial
links as well as the time link) 
and thus has the poorest signal.

\section{Quenched pion spectrum with unimproved staggered fermions}
\label{sec:pion:unimp}

\begin{table}[h!]
\begin{ruledtabular}
\begin{tabular}{ c c }
parameter & value \\
\hline
gauge action  & Wilson plaquette action \\
\# of sea quarks &  0 (quenched QCD) \\
$\beta$  & 6.0 \\
$1/a$ & 1.95 GeV (set by $\rho$ meson mass) \\
geometry & $16^3 \times 64$  \\
\# of confs & 370  \\
gauge fixing & Coulomb  \\
bare quark mass ($a m_q$) & 0.005, 0.01, 0.015, 0.02, 0.025, 0.03  \\
$Z_m$ & $\approx 2.5$  \\
\end{tabular}
\end{ruledtabular}
\caption{Simulation parameters for quenched study of
unimproved staggered fermions.}
\label{tab:par:unimp}
\end{table}

We begin with our results for unimproved staggered fermions
on quenched lattices. 
Although, as described in the introduction,
unimproved staggered fermions are not useful
for most practical applications, we begin with them because
they provide the benchmark against which improvements can be measured.
Furthermore, we can test our methods of calculating the pion
spectrum by comparing some of our results to those in the literature.
A final motivation is that we are able to add new information by
calculating with a wider range of quark masses than used previously.

The simulation parameters are given in Table~\ref{tab:par:unimp}.
Here $Z_m$ is
the factor by which the bare quark masses must be multiplied to obtain
the renormalized mass in the $\overline{\rm MS}$ scheme at a scale of
$2\;$GeV. We have only calculated the
spectrum for equal quark and anti-quark masses, and only consider
flavor non-singlet pions (so that there are no quark-disconnected
contractions). To set the scale, we note that $am=0.025$ corresponds
approximately to the physical kaon mass.

For $am=0.01$ and $0.02$, we can compare our results to those of
Ref.~\cite{ref:tsukuba}, where the complete pion spectrum was
calculated using similar methods.\footnote{%
Our results are also consistent with those of
Refs.~\cite{ref:jlqcd:1,ref:GGKS} but we do not show a detailed
comparison since both of these works used Landau gauge for which the
theoretical interpretation of the spectrum is not fully justified.}
%
\begin{table*}
\begin{ruledtabular}
\begin{tabular}{ c  c  c  c  c }
    &  \multicolumn{2}{c}{Ref.~\protect\cite{ref:tsukuba}} & \multicolumn{2}{c}{This work} \\
Operator & $am = 0.01$ & $am = 0.02$  
         & $am = 0.01$ & $am = 0.02$ \\
\hline
$(\gamma_5 \otimes \xi_5)$ &
0.23877(98) & 0.33476(82) &
0.2430(11)  & 0.3382(9) \\
$(\gamma_5 \otimes \xi_{i5})$ &
0.2896(14) & 0.3800(12)  &
0.2937(25) & 0.3835(16) \\
$(\gamma_5 \otimes \xi_{i4})$ &
0.3056(16) & 0.3933(14)  &
0.3104(27) & 0.3967(21) \\
$(\gamma_5 \otimes \xi_4)$ &
0.3167(20) & 0.4016(16) &
0.3230(48) & 0.4061(28) \\
\hline
$(\gamma_5 \otimes I)$ &
0.3300(46) & 0.4110(26) &
0.3323(57) & 0.4127(40) \\
$(\gamma_5 \otimes \xi_{45})$ &
0.2922(30) & 0.3826(18)  &
0.2978(62) & 0.3866(35) \\
$(\gamma_5 \otimes \xi_{ij})$ &
0.3085(26) & 0.3957(18)  &
0.3108(46) & 0.3980(33) \\
$(\gamma_5 \otimes \xi_{i})$ &
0.3168(33) & 0.4026(19) &
0.3211(49) & 0.4061(38) \\
\end{tabular}
\end{ruledtabular}
\caption{Comparison of results for $(am_\pi)$ with unimproved staggered
fermions with those from Ref.~\cite{ref:tsukuba}.  
Our results are those with
Golterman sink operators and cubic wall sources.}
\label{tab:cmp:unimp:2}
\end{table*}
The comparison is shown in Table \ref{tab:cmp:unimp:2}. 
Our results are systematically higher, although for most
pions the difference is only $1-2\sigma$. 
For the Goldstone, taste $\xi_5$ pion, however, the difference
is larger, $\sim 3-4\sigma$.
These differences could be due to the details of fitting
(our fitting range starts at somewhat later times), and they could
at least in part indicate a finite volume effect 
(our lattices are smaller: $16^3\times 64$ versus $20^3\times 40$). 
In any case, we consider the agreement good enough for the subsequent 
discussion of taste-breaking effects.

As a test of our methods, it is also interesting to compare the
relative size of our errors to those of Ref.~\cite{ref:tsukuba}.
The errors are comparable for  the Goldstone pion,
while ours grow more quickly with the distance between quark and
antiquark in the bilinears. We both use Coulomb-gauge sinks
(although time-directed links are treated differently), so
the differences between the
calculations are the number of configurations (370 for us, versus 50),
spatial volume (ours is half the size), and number and type of sources
(Ref.~\cite{ref:tsukuba} uses one wall-source per color, and thus three
times as many per configuration as we use).  We speculate that the
latter difference, and in particular the lack of noise from the U(1)
sources, is responsible for the smaller errors of Ref.~\cite{ref:tsukuba}.

\begin{figure}[t!]
\includegraphics[width=20pc]{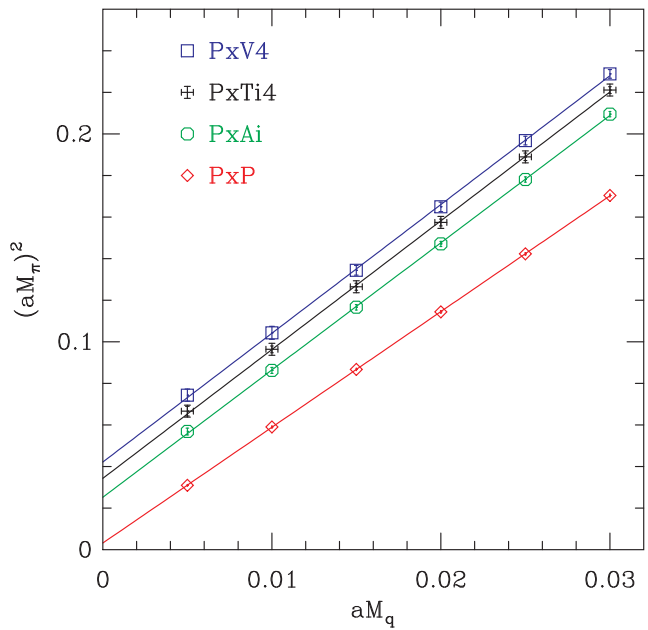}
\caption{$(a m_\pi)^2$ vs. $a m_q$ for unimproved quenched staggered
fermions, with cubic wall sources and Golterman sinks. Only LT tastes
are shown. Solid lines give the results of linear fits to the
quark-mass dependence.}
\label{fig:pion:unimp-gol-cw}
\end{figure}
\begin{figure}[t!]
\includegraphics[width=20pc]{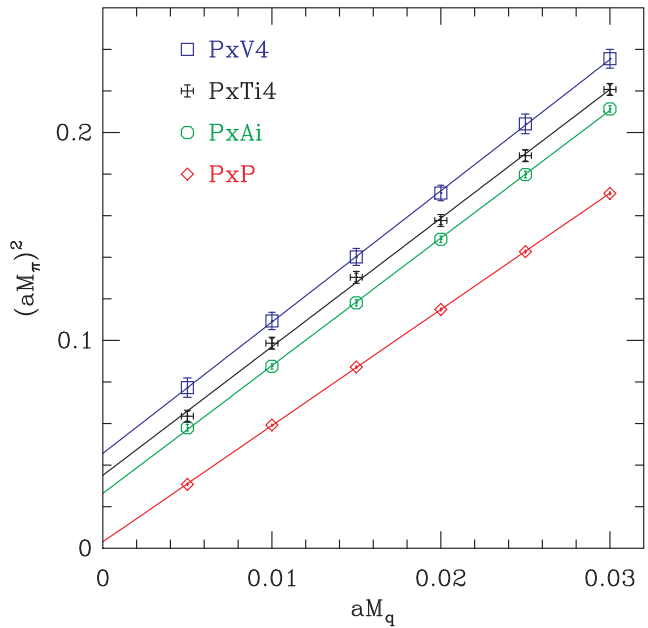}
\caption{As for Fig.~\protect\ref{fig:pion:unimp-gol-cw}
except using cubic U(1) sources.}
\label{fig:pion:unimp-gol-cu1}
\end{figure}

\begin{figure}[t!]
\includegraphics[width=20pc]{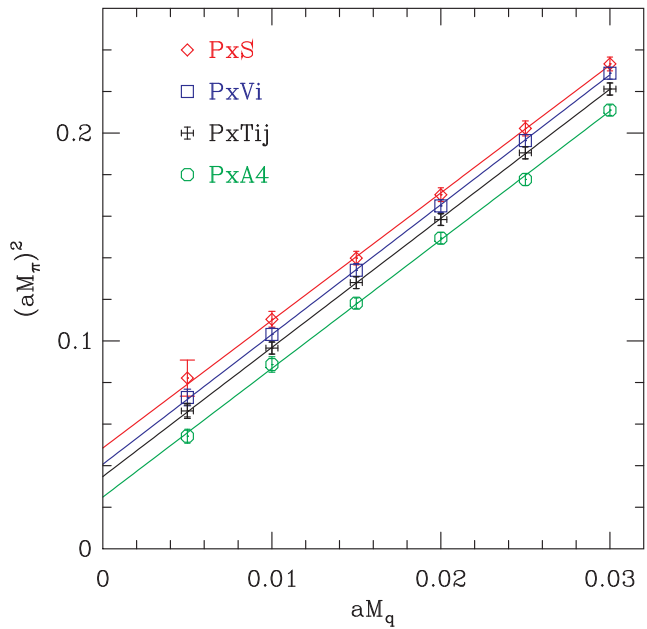}
\caption{$(a m_\pi)^2$ vs. $a m_q$ for NLT tastes
using  unimproved quenched staggered fermions, 
with cubic wall sources and Golterman sinks. 
Solid lines give the results of linear fits to the
quark-mass dependence.}
\label{fig:pion:unimp-gol-nlt-cw}
\end{figure}

\begin{figure}[t!]
\includegraphics[width=20pc]{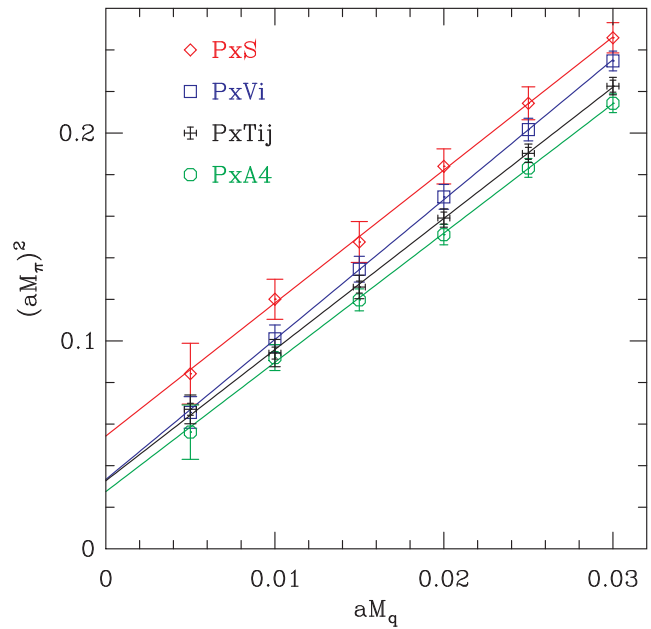}
\caption{ As for Fig.~\protect\ref{fig:pion:unimp-gol-nlt-cw}
except with cubic U(1) sources.}
\label{fig:pion:unimp-gol-nlt-cu1}
\end{figure}

We show the comparison of the two sources for LT pions
in Figs.~\ref{fig:pion:unimp-gol-cw}
 and \ref{fig:pion:unimp-gol-cu1},
and for NLT pions in 
Figs.~\ref{fig:pion:unimp-gol-nlt-cw} and
~\ref{fig:pion:unimp-gol-nlt-cu1}.\footnote{%
We do not focus here on the detailed chiral behavior, and thus use
linear fits. As the figures show, these give a good representation of
our data over the range of quark masses. We are not concerned that
the Goldstone pion mass does extrapolate exactly to zero at 
vanishing quark mass, since this can be explained by the
missing higher-order terms in the chiral expansion.}
In both cases the results from the two sources are consistent.
We note, however, that the cubic wall sources lead to noticeably smaller errors
for NLT pions than the cubic U(1) sources, although there is
no such difference for LT pions.

%

\begin{figure}[t!]
\includegraphics[width=20pc]{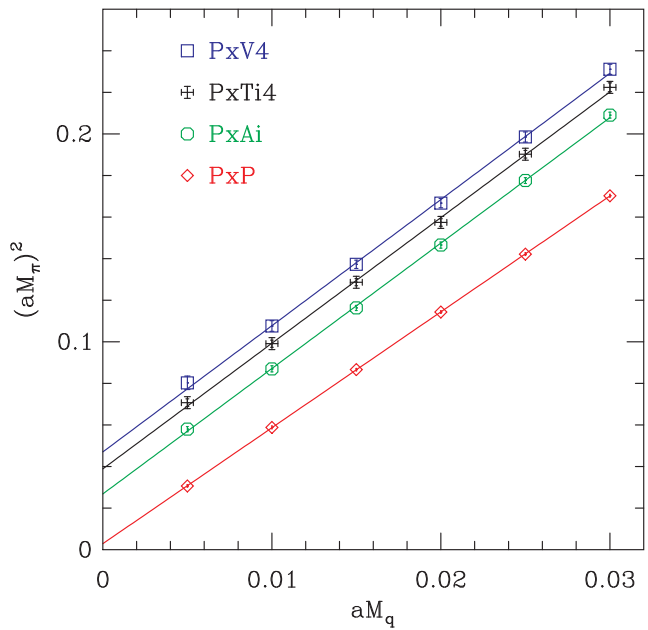}
\caption{$(a m_\pi)^2$ vs. $a m_q$ for LT tastes using
unimproved quenched staggered
fermions, with cubic wall sources and HPC sinks.}
\label{fig:pion:unimp-klu-cw}
\end{figure}

\begin{figure}[h!]
\includegraphics[width=20pc]{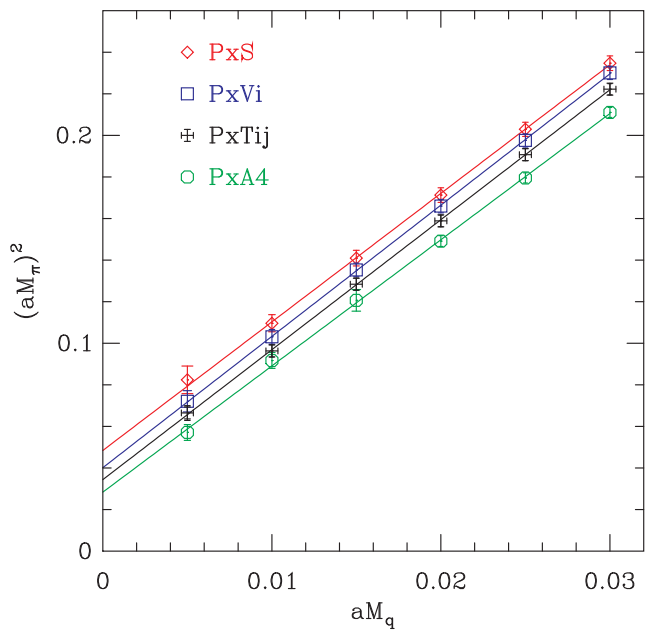}
\caption{As for Fig.~\protect\ref{fig:pion:unimp-klu-cw}
except for NLT pions.}
\label{fig:pion:unimp-klu-nlt-cw}
\end{figure}

We show results with HPC sinks (and cubic wall sources)
in Figs.~\ref{fig:pion:unimp-klu-cw} and
\ref{fig:pion:unimp-klu-nlt-cw}.
Comparing these with Figs.~\ref{fig:pion:unimp-gol-cw}
and \ref{fig:pion:unimp-gol-nlt-cw}, respectively,
we see that the choice of sinks has no significant impact
on the signal. (This is shown in more detail for
HYP fermions below.)
Thus we conclude that the extra irreps coupled to by the HPC
sources do not significantly degrade the signal.
Nevertheless, all else being equal, we prefer the Golterman
sources.

\bigskip

We now turn to the nature of the taste breaking with unimproved
fermions. Having results for all the tastes over a range of relatively
light quark masses (down to $\approx m_s^{\rm phys}/5$) allows us to
disentangle the different contributions.  We note that the splittings
are comparable to the pion mass-squareds themselves at the lightest
quark masses, indicating that the appropriate power-counting is ${\cal
O}(a^2) \approx {\cal O}(p^2)$. As seen in previous studies, the
states are ordered according to the number of gauge-links traversed
when moving between quark and anti-quark fields in the bilinear,
although the splittings are not directly proportional to this number.
In addition, we clearly observe that the slopes differ between the
tastes: the splittings all increase with $m_\pi^2$, with almost a
factor of two increase between the chiral limit and $am= am_s^{\rm
phys}\approx 0.025$. 
(The values of the slopes for LT tastes are given in
Table~\ref{tab:slope:hyp:unimp} below.)
This difference in slopes arises at ${\cal
O}(a^2 p^2)$ in chiral power-counting, and thus is a
next-to-leading-order (NLO) effect.  We thus expect it to be a small
correction for our lightest quark masses, increasing to a significant
correction for $m\approx m_s^{\rm phys}$.  This is indeed what we
find.

Staggered chiral perturbation theory predicts that the NLO
contributions to the slopes for the four different LT tastes are {\em
independent}~\cite{ref:RuthSS}.  In fact, the numerical results tell us
that the slopes for the heavier three tastes are consistent with each
other, while being significantly larger than that for the Goldstone
pion.  This implies that some of the coefficients in the NLO staggered
chiral Lagrangian are significantly smaller than others.

Another NLO prediction is that SO(4) taste symmetry should be broken,
i.e. that the leading order (LO) degeneracy between tastes $A_i$ and
$A_4$, $T_{i4}$ and $T_{ij}$, and $V_4$ and $V_i$, will no longer
hold. The splitting is caused by 
terms of ${\cal O}(a^2 p^2)$, and thus
affects only the slopes $(a m_\pi)^2/(a m_q)$, and not the
intercepts~\cite{ref:RuthSS}.  The difference in slopes {\em within}
SO(4) irreps is predicted to be of the same order as that {\em
between} SO(4) irreps. Taken at face value, this would imply that the
SO(4) symmetry would be completely broken at the point that NLO
contributions approach the size of those of LO (which, as we have
seen, occurs at the upper end of our mass range).

In fact, we find no such breakdown in the SO(4) symmetry.  The
accuracy to which this symmetry holds can be seen most easily from
Table~\ref{tab:cmp:unimp:2}, and also by comparing
Figs.~\ref{fig:pion:unimp-klu-cw} and \ref{fig:pion:unimp-klu-nlt-cw}
(the points with the same colors in the two figures live in the same
SO(4) multiplet).  The errors for NLT tastes, while larger than for
the LT tastes, are small enough that we can rule out a scenario with
complete breakdown of SO(4) symmetry for our largest quark masses.  In
fact, our results are consistent with SO(4) symmetry for all quark
masses.  Thus we conclude that our results are consistent with the
prediction of no SO(4) breaking in the intercepts, and that the
coefficients of the SO(4) breaking in the slopes must be much smaller
than the generic size.


\section{Quenched pion spectrum with HYP-smeared staggered fermions}
\label{sec:pion:hyp}

\begin{table}[h!]
\begin{ruledtabular}
\begin{tabular}{ c c }
parameter & value \\
\hline
gauge action  & Wilson plaquette action \\
\# of sea quarks &  0 (quenched QCD) \\
$\beta$  & 6.0 \\
$1/a$ & 1.95 GeV (set by $\rho$ meson mass) \\
geometry & $16^3 \times 64$  \\
\# of confs & 370  \\
gauge fixing & Coulomb  \\
smearing method & HYP (II) of Ref.~\cite{ref:wlee:2}  \\
bare quark mass ($a m_q$) & 0.01, 0.02, 0.03, 0.04, 0.05  \\
$Z_m$ & $\approx 1$
\end{tabular}
\end{ruledtabular}
\caption{Simulation parameters for the quenched study with
HYP-smeared staggered fermions.}
\label{tab:par:hyp}
\end{table}
%

\begin{table*}
\begin{ruledtabular}
\begin{tabular}{ c c c c c }
    &  \multicolumn{2}{c}{HPC} & \multicolumn{2}{c}{Golterman} \\
Operator & CU1 & CW & CU1 & CW \\
\hline
$(\gamma_5 \otimes \xi_5)$ & 
0.2262(27) & 0.2261(26) & 0.2263(27) & 0.2267(35) \\
$(\gamma_5 \otimes \xi_{i5})$ &
0.2313(27) & 0.2311(31) & 0.2314(30) & 0.2302(26) \\
$(\gamma_5 \otimes \xi_{i4})$ &
0.2355(29) & 0.2348(32) & 0.2356(29) & 0.2339(27) \\
$(\gamma_5 \otimes \xi_4)$ &
0.2395(30) & 0.2375(31) & 0.2396(31) & 0.2374(29) \\
\hline
$(\gamma_5 \otimes \xi_{45})$ & 
0.2293(74) & 0.2239(69) & 0.2343(72) & 0.2239(69) \\
$(\gamma_5 \otimes \xi_{ij})$ & 
0.2338(77) & 0.2264(56) & 0.2346(79) & 0.2233(61) \\
$(\gamma_5 \otimes \xi_{i})$ & 
0.2392(79) & 0.2281(56) & 0.2409(65) & 0.2262(58) \\
$(\gamma_5 \otimes I)$ & 
0.2407(83) & 0.2298(62) & 0.2403(84) & 0.2309(65) \\
\end{tabular}
\end{ruledtabular}
\caption{Source and sink dependence of $(am_\pi)$.
Results using HYP-smeared staggered fermions at $am=0.02$
on the quenched lattices.
HPC and Golterman denote the choice of sink,
while CU1 and CW denote cubic U(1) and cubic wall sources, respectively.}
\label{tab:cmp:hyp:3}
\end{table*}

Using the same set of quenched gauge configurations as in
Sec.~\ref{sec:pion:unimp}, we now study the pion spectrum with
HYP-smeared staggered fermions.  The parameters we
use are summarized in Table~\ref{tab:par:hyp}.  We recall that
HYP smearing is straightforward to implement in practice: one simply
applies HYP smearing to the gauge configuration and then calculates
the propagators using the unimproved staggered action on the resulting
lattice. HYP smearing is carried out after Coulomb gauge fixing.

Despite using larger bare quark masses, our calculations with HYP-smeared
fermions correspond to physical quarks which are
slightly lighter than those used for
unimproved staggered fermions in the previous section.  This is
because $Z_m \approx 1$ for HYP-smeared staggered fermions, whereas
$Z_m \approx 2.5$ for unimproved staggered fermions.
In particular, the physical strange quark mass is 
$a m_s^{\rm phys}\approx 0.052$, 
slightly larger than our largest quark mass~\cite{ref:wlee:BK},
whereas the physical strange quark mass for unimproved fermions
is approximately equal to the next-to-largest quark mass 
in that calculation ($a m_q=0.025$).

Table~\ref{tab:cmp:hyp:3} shows the dependence of
pion masses on source and sink type for one choice of quark mass.
As for unimproved quarks, the sink choice is unimportant
(the central values are consistent, and the errors 
are very similar).
It also remains true that the cubic wall sources lead to
smaller errors for the NLT pions, although the reduction in the
errors is a much smaller effect.
While we do not understand why this is, it does mean that,
in practice, the choice of source appears less crucial for
HYP-smeared staggered fermions.

\begin{figure*}[t!]
\includegraphics[width=20pc]{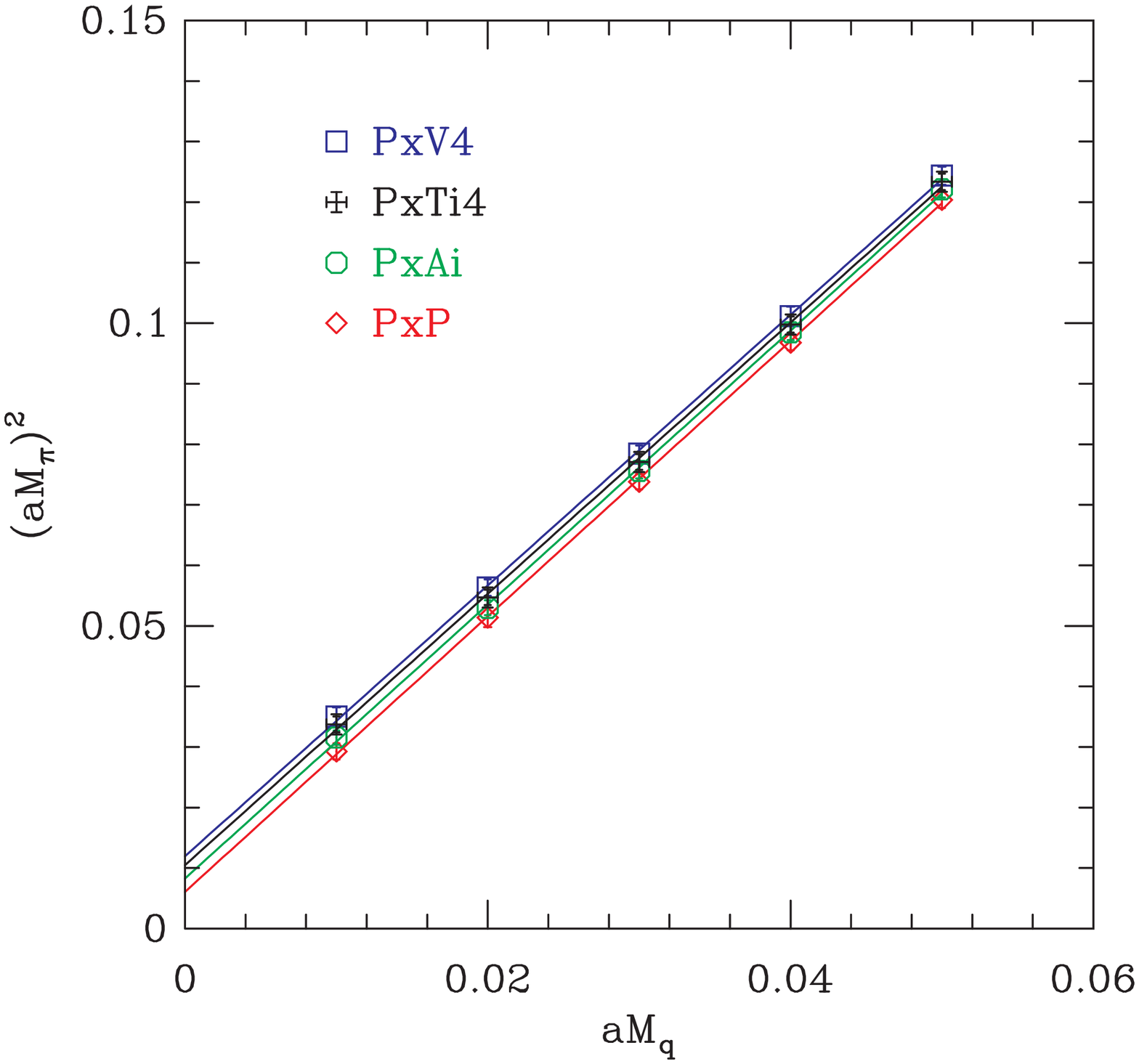}
\includegraphics[width=20pc]{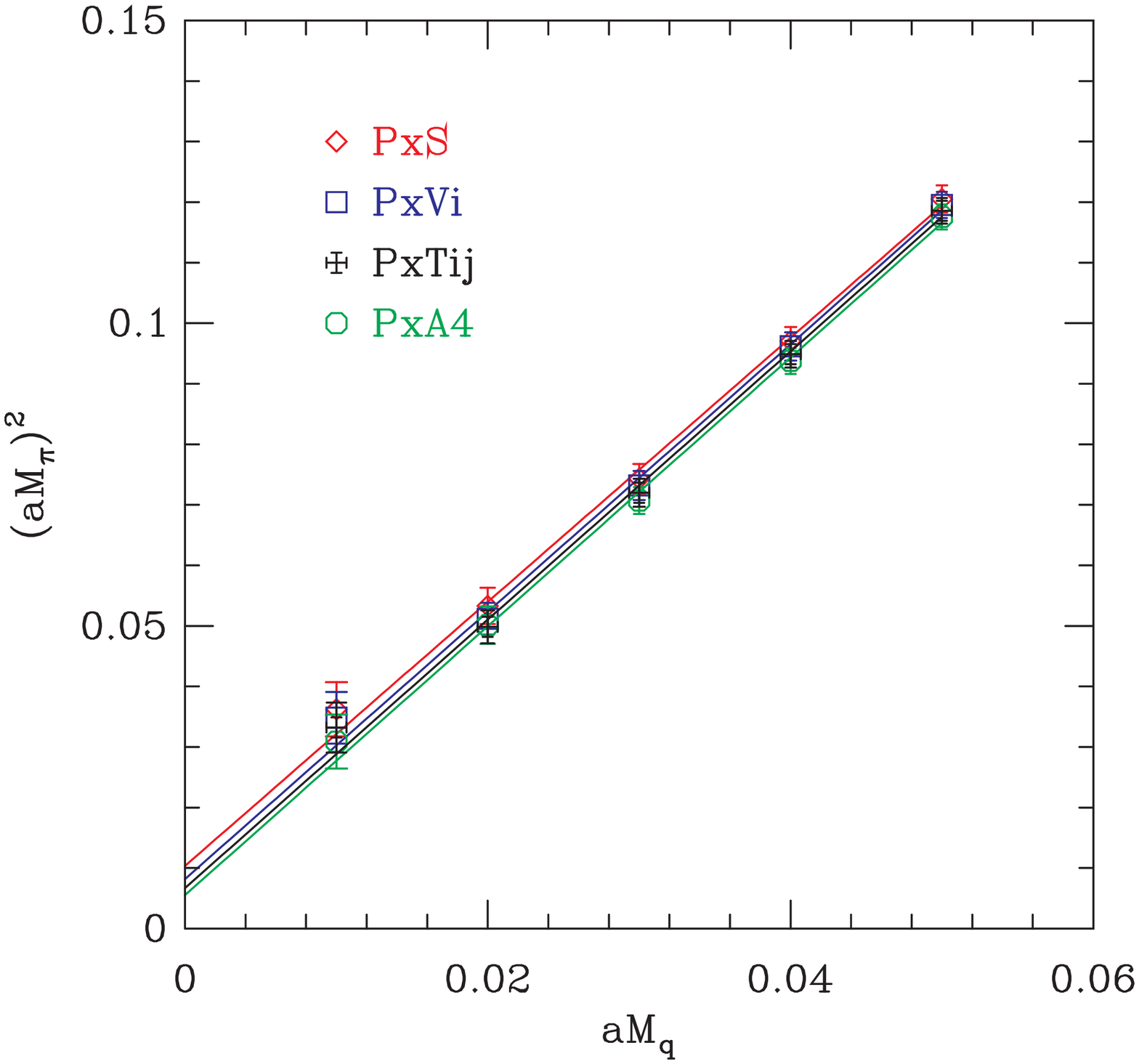}
\caption{$(a m_\pi)^2$ vs. $a m_q$ for HYP-smeared staggered fermions
in quenched QCD, with cubic wall sources and Golterman sinks. Linear
fits to the quark-mass dependence are shown. The left panel
shows LT tastes; the right panel shows NLT tastes.}
\label{fig:pion:hyp-gol-cu1}
\end{figure*}
%

\begin{table}[h!]
\begin{ruledtabular}
\begin{tabular}{c c c}
Taste ($F$) & $\Delta_F$ (unimproved) & $\Delta_F$ (HYP) \\
\hline
$\xi_{i 5}$ & 0.0233(14) &  0.0025(11)  \\
$\xi_{i 4}$ & 0.0320(27) &  0.0050(14)  \\
$\xi_{4}  $ & 0.0425(49) &  0.0073(13) \\
\hline
$\xi_{4 5}$ & 0.0244(91) &  0.0036(53)  \\
$\xi_{i j}$ & 0.0296(73) &  0.0050(51)  \\
$\xi_{i}  $ & 0.0303(78) &  0.0076(47) \\
$I        $ & 0.0511(147) &  0.0078(56) \\
\end{tabular}
\end{ruledtabular}
\caption{Comparison of taste symmetry breaking 
between unimproved and HYP-smeared fermions. 
Here $\Delta_F = [a m_\pi(F)]^2 - [a m_\pi(\xi_5)]^2$,
extrapolated to the chiral limit.
Both results are obtained using Golterman sinks
and cubic U(1) sources in quenched QCD at $\beta=6.0$.}
\label{tab:delta:hyp:unimp}
\end{table}

The resulting pion mass-squareds are shown for our preferred source and sink,
and for all tastes, in Fig.~\ref{fig:pion:hyp-gol-cu1}.
Comparing to the previous figures, we see that taste-breaking
is substantially reduced.\footnote{%
The vertical scales in the plots for the
unimproved and HYP action differ because the quarks,
and thus the pions, are lighter with the latter.}
A quantitative measure of this is given in
Table~\ref{tab:delta:hyp:unimp}, which collects the mass-squared
splittings after linear extrapolation to the chiral limit,
both for the HYP and the unimproved action.
Note that this measure is insensitive to the particular
range of quark masses chosen (as long as they are small enough
that linear fits work well).
For the LT tastes, where the splittings with the HYP action
are statistically significant, 
we find that HYP smearing reduces the splittings
by a factor of $5 -10$.
The ordering of states is unchanged.
For the NLT tastes,
the splittings are reduced by similar factors,
but the errors are large enough that a more quantitative statement
cannot be made.
For the same reason, we cannot make any quantitative comments
on the size of possible SO(4) splitting.

Our results for the reduction in taste-breaking
are qualitatively consistent with those obtained
previously~\cite{ref:hyp:1,ref:orginos:1,ref:mason:1} 
(although differences in simulation details
do not allow a quantitative comparison).
Having the results for a range of quark masses allows
us to draw some conclusions on the appropriate power-counting.
In particular, given the smallness of the splittings, it 
may be appropriate for this range of quark masses to treat the 
these ${\cal O}(a^2)$ effects as of NLO rather than of LO,
i.e. to assume that $a^2 \sim p^4 \sim m_q^2$.

\begin{table}[h!]
\begin{ruledtabular}
\begin{tabular}{c c c}
Taste ($F$) & $c_2$ (unimproved) & $c_2$ (HYP) \\
\hline
$\xi_{5}$   & 5.57(2) &  2.28(4)  \\
$\xi_{i 5}$ & 6.08(4) &  2.26(4)  \\
$\xi_{i 4}$ & 6.03(9) &  2.24(4)  \\
$\xi_{4}  $ & 5.99(10) &  2.24(4) \\
%
\end{tabular}
\end{ruledtabular}
\caption{Comparison of taste symmetry breaking in slopes
between unimproved
and HYP-smeared fermions. 
$c_2$ is defined by fitting the data to $ [a m_\pi(F)]^2 = c_1 + c_2 (a m_q)$.
Both results are obtained using Golterman sinks
and cubic wall sources in quenched QCD at $\beta=6.0$.}
\label{tab:slope:hyp:unimp}
\end{table}

A test of this idea is provided by the taste-dependence of the slopes
in the $m_\pi^2$ vs. $m_q$ plots.  As noted in the previous section,
this difference is an ${\cal O}(a^2 p^2)$ effect, and thus of NLO in
the standard power counting of staggered chiral perturbation theory.
It is observed at the expected order of magnitude with unimproved
staggered fermions (see, e.g., Fig.~\ref{fig:pion:unimp-gol-cw}).  If
the power-counting changes to $a^2\sim p^4$ for HYP-smeared fermions
then the difference in the slopes would be of NNLO (next to next to
leading order), and thus expected to be very small for our range of
quark masses.  This is indeed what we find.
Table~\ref{tab:slope:hyp:unimp} collects the slopes for the LT tastes,
and we see that, with HYP-smeared fermions, they are equal for all
tastes within the statistical uncertainty, which is at the 2\% level.

\section{Comparison of asqtad and HYP-smeared staggered fermions}
\label{sec:cmp:asqtad-hyp}

\begin{table}[h!]
\begin{ruledtabular}
\begin{tabular}{ c  c }
parameter & value \\
\hline
gauge action  & 1-loop tadpole-improved Symanzik \\
sea quarks &  $2+1$ flavors of asqtad staggered \\
sea quark masses & $am_\ell=0.01$, $am_s=0.05$ \\
$\beta$  & 6.76 \\
$a$ & 0.125 fm \\
geometry & $20^3 \times 64$  \\
\# of confs & 640 (asqtad)/ 671 (HYP) \\
gauge fixing & Coulomb  \\
valence quark type & asqtad and HYP staggered \\
valence quark mass (asqtad) & 0.007, 0.01, 0.02, $\ldots$, 0.05  \\
valence quark mass (HYP) & 0.005, 0.01, 0.015, $\ldots$, 0.05  \\
$Z_m$ & $\approx 1$
\end{tabular}
\end{ruledtabular}
\caption{Simulation parameters used for the comparison
of asqtad and HYP-smeared valence quarks.
The HYP smearing uses
the parameters of case (II) in Ref.~\cite{ref:wlee:2}.}
\label{tab:par:unquenched}
\end{table}

In this section we compare two types of improved staggered quark
actions: the ``asqtad'' action used by the MILC
collaboration and the HYP-smeared action.  
We use one of the so-called ``coarse'' lattice ensembles generated by
the MILC collaboration~\cite{ref:coarseMILC}, with the average light
quark mass roughly one fifth of the physical strange quark mass.  
The parameters of the study are summarized in
Table~\ref{tab:par:unquenched}.  We note that the conventions
of the asqtad action are such that, when comparing to the HYP
action, the appropriate bare quark mass to use is $a m_q/u_0$,
where $u_0$ is the ``average link''~\cite{ref:milc:1}.
In other words, if the HYP and asqtad quark masses are nominally equally,
the HYP masses are actually smaller by a
factor of $u_0 = 0.86774$ (this value coming from Ref.~\cite{ref:milc:1}
using one of the conventional definitions of $u_0$).
To set the scale, the strange quark mass with the asqtad action
is $\approx 0.045$, i.e. near the upper end of the range used,
while that for the HYP action is 
$\approx 0.057$, which is somewhat above our heaviest quark mass.
The results for asqtad valence quarks are from
Ref.~\cite{ref:milc:1}, while those for HYP-smeared quarks are new.

We only have results for HYP-smeared quarks using the cubic U(1)
source.  Although the results of the previous section indicate that
this source is slightly inferior for NLT pions, our errors are small
enough for the purposes of this study.

When we calculate HYP-smeared staggered quark propagators using the
cubic U(1) source, we choose the source time slice randomly
configuration by configuration such that unwanted autocorrelations
in the two-point pion correlation functions are removed.

\begin{figure}[t!]
\includegraphics[width=20pc]{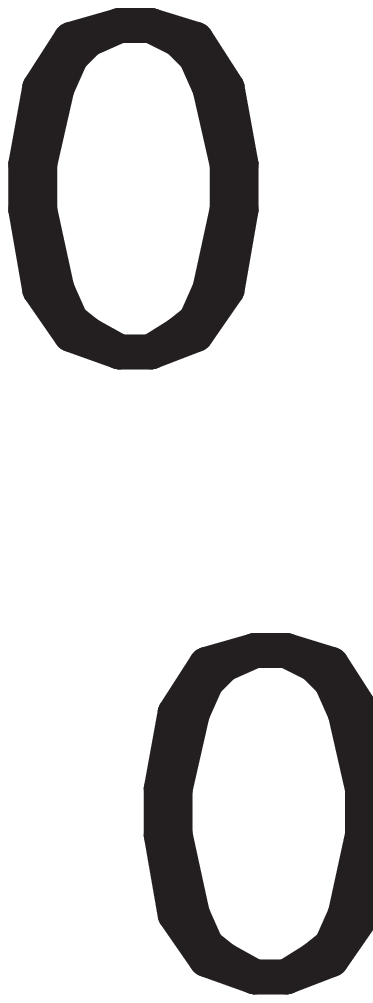}
\caption{$(r_1 m_\pi)^2$ vs. $2 r_1 m_q$ on unquenched configurations,
with asqtad valence quarks. All tastes are shown. The correspondence of
the taste labels to those in other figures is as follows:
$\pi_5\equiv\,$PxP, $\pi_{05}\equiv\,$PxA4, $\pi_{i5}\equiv\,$PxAi,
$\pi_{0i}=\equiv\,$PxTi4, $\pi_{ij}\equiv\,$PxTij
$\pi_0\equiv\,$PxV4, $\pi_i\equiv\,$PxVi, and
$\pi\equiv\,$PxS. Results are from Ref.~\protect\cite{ref:milc:1}.
The lines are linear fits to each taste.}
\label{fig:pion-spec-milc}
\end{figure}

\begin{figure}[t!]
\includegraphics[width=20pc]{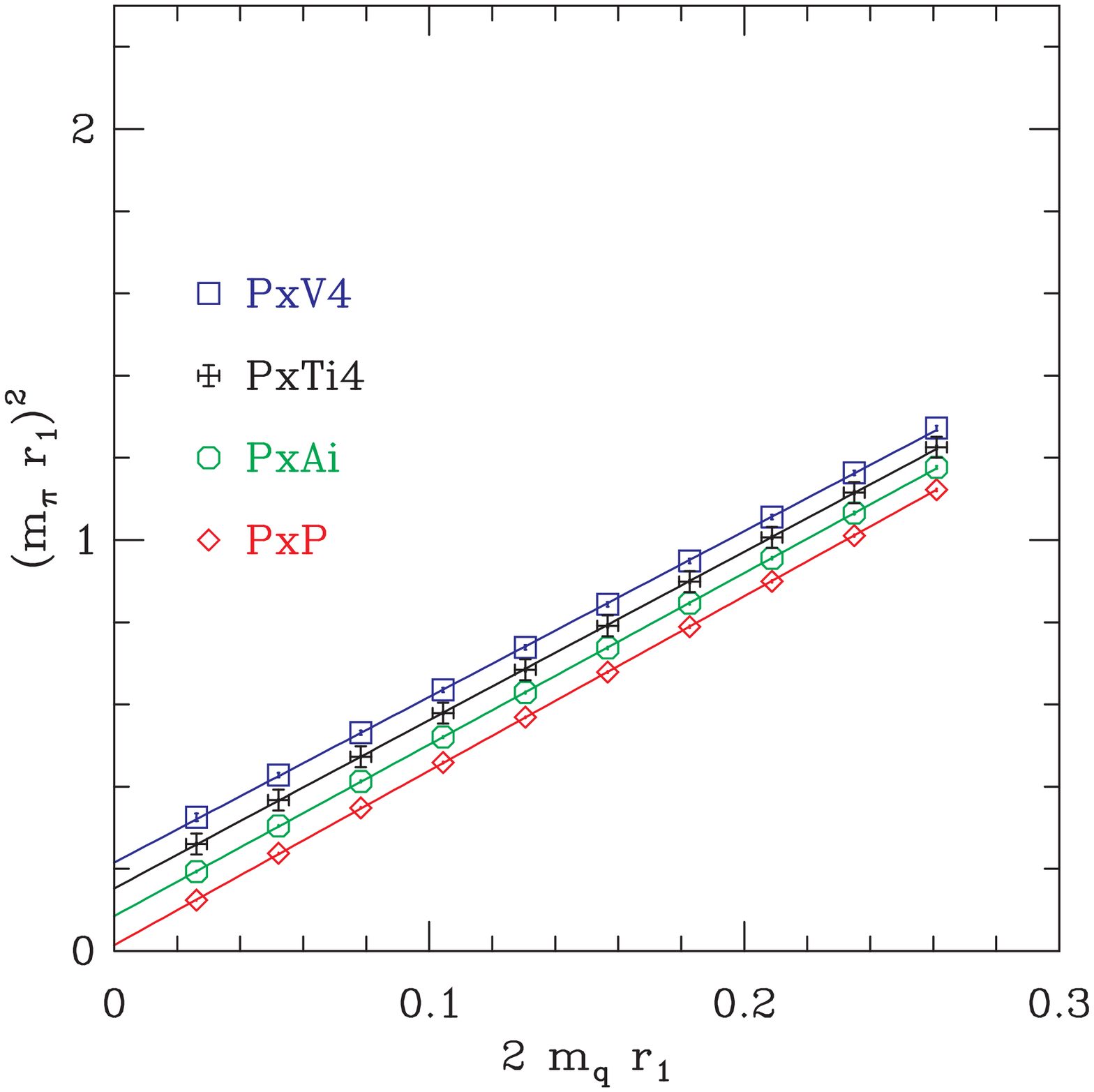}
\caption{As for Fig.~\protect\ref{fig:pion-spec-milc} for LT tastes
except using HYP-smeared valence staggered fermions, with Golterman
sinks and cubic U(1) sources.}
\label{fig:pion:mixed-gol-cu1}
\end{figure}

\begin{figure}[t!]
\includegraphics[width=20pc]{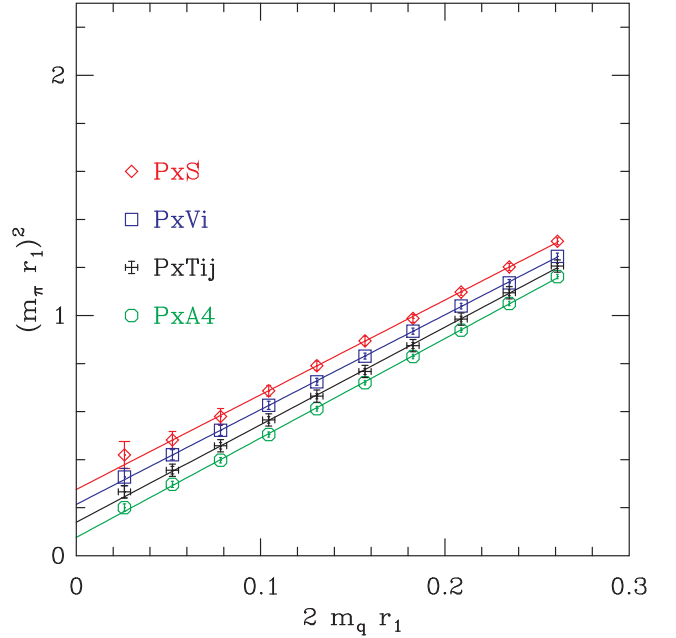}
\caption{As for Fig.~\protect\ref{fig:pion-spec-milc} for NLT tastes
except using HYP-smeared valence staggered fermions, with Golterman
sinks and cubic U(1) sources.}
\label{fig:pion:mixed-gol-nlt-cu1}
\end{figure}

In Fig.~\ref{fig:pion-spec-milc}, we show the results from
Ref.~\cite{ref:milc:1} for $(m_\pi)^2$ for all tastes as a function of
quark mass for asqtad valence quarks.  This should be compared to our
results with HYP-smeared valence fermions, given in
Fig.~\ref{fig:pion:mixed-gol-cu1} and
Fig.~\ref{fig:pion:mixed-gol-nlt-cu1}. Note that different axes are
used from the earlier plots, with the scale here being set by $r_1$,
which is $2.61 a$ on these lattices.


Comparing the two valence actions, we see that the taste-breaking is
substantially reduced for HYP-smeared fermions. To make this
quantitative we quote the values for some of the taste splittings,
extrapolated to the chiral limit using linear fits, in
Table~\ref{tab:delta:unquenched}. The splittings with HYP smearing are
smaller by a factor of $2.5-3$. This is consistent qualitatively
with the result noted in the introduction from Ref.~\cite{ref:HISQ}
obtained on quenched lattices. It is also comparable to the 
factor of $3.6(5)$ improvement obtained using HISQ rather
than asqtad fermions on similar dynamical lattices~\cite{ref:HISQ}.

\begin{table}[t!]
\begin{ruledtabular}
\begin{tabular}{ccc}
Taste ($F$) & $\Delta_F$ (asqtad) & $\Delta_F$ (HYP) \\
\hline
$\xi_{i 5}/\xi_{4 5}$  & 0.205(2) &  0.0708(12)/0.062(13)  \\
$\xi_{i 4}/\xi_{i j}$  & 0.327(4) &  0.1380(29)/0.125(16)  \\
$\xi_{4}/\xi_{i} $     & 0.439(5) &  0.2008(62)/0.199(26) \\
$I         $           & 0.537(15) & 0.257(34) \\
\end{tabular}
\end{ruledtabular}
\caption{Taste symmetry breaking effect: 
$\Delta_F = [r_1 m_\pi(F)]^2 - [r_1 m_\pi(\xi_5)]^2$
extrapolated to the chiral limit. Note that the scale
is set here by $r_1$ and not the $1/a$ used earlier.
Results from unquenched lattices comparing valence
asqtad (from Ref.~\cite{ref:milc:1}) with HYP-smeared
fermions (Golterman sinks with cubic U(1) sources).}
\label{tab:delta:unquenched}
\end{table}

A surprising feature of the comparison of Figs.~\ref{fig:pion-spec-milc}
with Figs.~\ref{fig:pion:mixed-gol-cu1} and
\ref{fig:pion:mixed-gol-nlt-cu1}
is that the slopes for the Goldstone pions differ: 
that for HYP-smeared fermions is about 1.4 times smaller
than that for asqtad fermions. 
(The difference is emphasized by our use of the same vertical scales.)
This slope is a parameter in the corresponding chiral Lagrangian,
and should agree for the two actions up to differences
in renormalization factors for the quark masses
(aside from taste-conserving scaling violations).
In fact, as already noted, the asqtad masses should be multiplied by
$1/u_0\approx 1.15$, 
so that the horizontal
scale in Fig.~\ref{fig:pion-spec-milc}
should be stretched by a factor of 1.15,
thus reducing the slope by this factor.
This leaves a factor of about 1.2 difference in slopes to explain.
This could be entirely due to differences in $Z_m$---
the two-loop result 
for asqtad fermions is $Z_m\approx 1.16$~\cite{ref:mason},
while that for the HYP action on asqtad sea-quarks is not known
yet at one-loop.
It is important to note, however, that the splittings given in
Table~\ref{tab:delta:unquenched} are in the chiral limit and thus
are unaffected by any relative factors between the
bare quark masses.

Returning to taste-breaking, we note that for both types of valence
fermion the slopes for different tastes are nearly equal.  For asqtad
fermions we also see that no SO(4) breaking is apparent even at the
largest quark masses (which, we recall, are close to the physical
strange quark mass).  Thus, despite the relatively large splittings
with asqtad fermions [of ${\cal O}(a^2)$ in the chiral counting],
which are clearly to be treated as comparable to the contributions
from the quark mass [of ${\cal O}(p^2)$], there are no indications of
NLO contributions [of ${\cal O}(a^2 p^2)$].  This is perhaps
surprising and is in any case in contrast to unimproved staggered
fermions where ${\cal O}(a^2 p^2)$ effects were clearly visible for a
similar range of quark masses.  For HYP-smeared fermions there is also
no indication of NLO effects. The slopes are very close and,
as can be seen from Table~\ref{tab:delta:unquenched}, 
there is no significant SO(4) breaking.

\begin{table*}
\begin{ruledtabular}
\begin{tabular}{ c c c c c }
    &  \multicolumn{2}{c}{quenched ($a\approx 0.1\;$fm)} & 
  \multicolumn{2}{c}{unquenched ($a\approx 0.125\;$fm)} \\ 
Taste($F$) & $\Delta_F$(unimproved) & $\Delta_F$(HYP)
         & $\Delta_F$(asqtad)     & $\Delta_F$(HYP) \\
\hline
$\xi_{\mu 5}$ &
0.0931(56) & 0.0094(41) & 0.0759(7) & 0.0262(5) \\
$\xi_{\mu \nu}$ &
0.128(11) & 0.0189(53) & 0.1211(14) & 0.0511(11) \\
$\xi_\mu$ &
0.170(20) & 0.0279(49) & 0.1625(19) & 0.0743(23) \\
\end{tabular}
\end{ruledtabular}
\caption{Comparison of taste-breaking for LT tastes on all lattices
considered in this paper. $\Delta_F = [m_\pi(F)]^2 - [m_\pi(\xi_5)]^2$,
now in units of ${\rm GeV}^2$.}
\label{tab:cmp:all}
\end{table*}

We next compare taste-breaking for HYP-smeared
fermions between the quenched lattices considered earlier and the
dynamical lattices. Clearly, the splittings are significantly larger
for the latter. To make this quantitative we compare the
splittings in physical units in Table~\ref{tab:cmp:all}, where
we see that they are 2-2.5 larger on the unquenched than on the quenched
lattices with HYP-smeared fermions. Can we understand
this?  The answer is yes, at least qualitatively.  Taste-breaking is
expected to scale as $\alpha^2 a^2$, since the contributions from
one-gluon exchange have been removed.  The lattice spacing on the
dynamical lattices is larger, with the ratio of $a^2$ to that on the
quenched lattices being about 1.5.  We also know that $\alpha$ at a
scale of $\approx 1/a$ (which is the scale relevant for taste-breaking
gluon vertices) is larger on the dynamical lattices, since the
long-distance physics has been (approximately) matched by
construction, and the coupling runs toward zero more slowly for the
dynamical lattices.  Combining these effects, we expect to find that
the taste-breaking for HYP-smeared fermions is perhaps twice as large
on the dynamical as on the quenched lattices.  This is 
in accord with our results.
Note, however, that $\alpha^2 a^2$ is reduced by about a factor
of 2.5 between the MILC coarse and fine lattices~\cite{ref:milc:1}.
Thus we expect the taste-splitting on the MILC fine lattices
with the HYP-action to be similar to that we observed above
on the $a=1/10\;$fm quenched lattices.

In summary, the main difference between asqtad and
HYP-smeared valence quarks is a substantial reduction in the LO
taste-breaking splittings between SO(4) multiplets. Unlike the
quenched case studied above, however, the power-counting
$a^2 \sim p^2 $ appears to remain necessary on these 
coarse MILC lattices.
It seems likely, however, that taste-breaking will be a NLO effect
with HYP-smeared quarks on the fine MILC lattices.

\section{Conclusion}
\label{sec:conclude}

In this paper, we have compared three versions of staggered 
fermions---unimproved, HYP-smeared, and asqtad---from the 
standpoint of taste-symmetry breaking.
We confirm and make quantitative
the hierarchy expected from previous work, namely that
HYP smearing is significantly more effective at reducing
taste-breaking than asqtad improvement.
This is summarized in Table~\ref{tab:cmp:all}.
Our results imply that HYP-smeared valence fermions on the fine MILC lattices
(with $a\approx 1/11\;$fm) should have taste-breaking small enough that it
will be possible to treat discretization errors as a NLO effect in chiral
perturbation theory (rather than as a LO effect as required with asqtad fermions,
and with HYP-smeared fermions on the coarse MILC lattices).
This is an encouraging result for our ongoing calculations of electroweak
matrix elements using HYP-smeared valence fermions.

To put the results into context, it is useful to convert the
size of the splittings into a scale characterizing discretization errors.
Taking the splitting of the 2-link LT pion (taste $\xi_{\mu\nu}$) as indicative
of an ``average'' splitting, and setting $\Delta m_\pi^2 = a^2 \Lambda^4$,
we find $\Lambda=0.5$ and $0.6\;$MeV for HYP fermions
on the quenched and unquenched lattices, 
respectively. These are scales within the range of sizes
that one would expect for generic discretization errors. In other words,
the improvement due to HYP smearing  brings the taste-breaking errors down into
the range expected of generic discretization errors, so that
discretization errors with staggered fermion are no longer anomalously 
large.

\section{Acknowledgments}

Helpful discussions with A.~Hasenfratz are acknowledged with gratitude.
C.~Jung is supported by the U.S.~Dept.~of Energy under contract
DE-AC02-98CH10886.
W.~Lee acknowledges with gratitude that the research at Seoul National
University is supported by the KICOS grant
K20711000014-07A0100-01410, by the KRF grants (KRF-2006-312-C00497 and
KRF-2007-313-C00146), and by the BK21 program of Seoul National
University. 
The work of S.~Sharpe is supported in part by the US Dept.~of Energy grant
DE-FG02-96ER40956. 
Computations for this work were carried out in part on
facilities of the USQCD Collaboration, which are funded by the
Office of Science of the U.S.~Department of Energy.

\appendix
\section{Anatomy of the hypercube operators}
\label{app:anatomy}
In Sec.~\ref{sec:op}, we defined the hypercube operators,
eq.~(\ref{eq:HPCop}),
and noted that they are not irreps of the time-slice group
of Ref.~\cite{ref:golterman:1}.
Here, following the procedure of Ref.~\cite{ref:sharpe:2}, we express
them in terms of true irreps.
For simplicity, we work in the free theory, i.e. we do not explicitly
include gauge links, but it is straightforward to add these at the
end, following, say, the prescription in the text.
Consider a single hypercube operator of general spin and taste:
\begin{equation}
O^{\rm Klu}_{S,F} = \sum_{A,B} \bar{\chi}(A) 
( \overline{\gamma_S \otimes \xi_F} )_{AB} \chi(B)
\label{eq:app:klu:op}
\,.
\end{equation}
Summing this over a timeslice as in eq.~(\ref{eq:HPCop}) leads to the
HPC operators we actually use, but we delay this summation
until necessary.
To simplify the discussion, we first assume that
$|S_\mu -F_\mu| = 1$ for one index $\mu$, while $|S_\nu -F_\nu| = 0$
for $\nu \ne \mu$. In other words, we
first consider a ``one-link'' bilinear.
We also need the ``averaging'' operator $\Phi_\mu$ given
in eq.~(\ref{eq:Phimudef}) and the derivative operator
\begin{eqnarray}
D_\mu \chi(A) &=& \frac{1}{2} [ \chi(A+\hat\mu) - \chi(A-\hat\mu)]
\,,
\end{eqnarray}
in terms of which
\begin{eqnarray}
\chi(A+\hat\mu) &=& \Phi_\mu \chi(A) + D_\mu \chi(A)
\\
\chi(A-\hat\mu) &=& \Phi_\mu \chi(A) - D_\mu \chi(A)
\,.
\end{eqnarray}

Now we can express the single-hypercube operator as
(summation over $A, B$ now implicit)%
\begin{eqnarray*}
O^{\rm Klu}_{S,F} &=& \bar{\chi}(A) 
( \overline{\gamma_S \otimes \xi_F} )_{AB}
(\delta_{B_\mu, A_\mu+1} + \delta_{B_\mu, A_\mu-1}) 
\chi(B)
\\ &=& \frac{1}{2} \bar{\chi}(A)
( \overline{\gamma_S \otimes \xi_F} )_{AB}
\delta_{B_\mu, A_\mu+1}
[ \Phi_\mu \chi(A) + D_\mu \chi(A) ]
\\ &+& \frac{1}{2} 
[ \Phi_\mu \bar{\chi}(B) - D_\mu \bar{\chi}(B)]
\delta_{A_\mu, B_\mu-1}
( \overline{\gamma_S \otimes \xi_F} )_{AB}
\chi(B)
\\ &+& \frac{1}{2}
\bar{\chi}(A)
( \overline{\gamma_S \otimes \xi_F} )_{AB}
\delta_{B_\mu, A_\mu-1}
[ \Phi_\mu \chi(A) - D_\mu \chi(A) ]
\\ &+& \frac{1}{2}
[ \Phi_\mu \bar{\chi}(B) + D_\mu \bar{\chi}(B)]
\delta_{A_\mu, B_\mu+1}
( \overline{\gamma_S \otimes \xi_F} )_{AB}
\chi(B)
\\ &=& \frac{1}{2} \bar{\chi}(A) 
( \overline{\gamma_S \otimes \xi_F} )_{AB} 
[ \Phi_\mu \chi(A) ]
\\ &+& \frac{1}{2} [ \Phi_\mu \bar{\chi}(B) ]
( \overline{\gamma_S \otimes \xi_F} )_{AB}
\chi(B)
\\ &+& \frac{1}{2} \bar{\chi}(A)
( \overline{\gamma_{\mu 5 S} \otimes \xi_{\mu 5 F}} )_{AB}
[ D_\mu \chi(A) ]
\\ &-& \frac{1}{2} [ D_\mu \bar{\chi}(B) ]
( \overline{\gamma_{\mu 5 S} \otimes \xi_{\mu 5 F}} )_{AB}
\chi(B)
\end{eqnarray*}
In the last step we have used the following
result for spin-taste matrices, assuming $|S_\mu-F_\mu|=1$:
\begin{equation}
( \overline{\gamma_S \otimes \xi_F} )_{AB} (B-A)_\mu
= ( \overline{\gamma_{\mu 5 S} \otimes \xi_{\mu 5 F}} )_{AB}
\,.
\end{equation}

Now we reinstate the sum over spatial hypercubes. 
Then the following identity holds:
\begin{eqnarray}
& & \sum_{\vec n} \bar{\chi}(A)
( \overline{\gamma_S \otimes \xi_F} )_{AB}
[ \Phi_\mu \chi(A) ]
\nonumber \\
& & = \sum_{\vec n} [ \Phi_\mu \bar{\chi}(B) ]
( \overline{\gamma_S \otimes \xi_F} )_{AB}
\chi(B)
\nonumber \\
& & \equiv  [\sum_{\vec n} \bar{\chi}
( \overline{\gamma_S \otimes \xi_F} )
\chi]_T
\label{eq:truerep}
\end{eqnarray}
where $[ \cdots ]_T$ indicates that this is part of
an irrep of the timeslice group of Ref.~\cite{ref:golterman:1}.
In fact, if $\mu$ is a spatial index, the $[ \cdots ]_T$ part of the HPC operator
is identical to the corresponding Golterman operator 
of eq.~(\ref{eq:Goltdef}). If $\mu=4$, the operator
in (\ref{eq:truerep}) transforms in the same irrep as the corresponding
Golterman operator, although it is not identical.

To pick out the remaining parts of the HPC operator,
we define, as in Ref.~\cite{ref:sharpe:2}
\begin{eqnarray}
& & \sum_{\vec n} 
\bar{\chi}
( \overline{\gamma_{\mu 5 S} \otimes \xi_{\mu 5 F}} )
\overleftrightarrow{D}_\mu \chi
\nonumber \\ & & 
\equiv \sum_{\vec n} \big\{ \bar{\chi}(A)
( \overline{\gamma_{\mu 5 S} \otimes \xi_{\mu 5 F}} )_{AB}
[ D_\mu \chi(A) ]
\nonumber \\ & & \hspace*{1pc}
- [ D_\mu \bar{\chi}(B) ]
( \overline{\gamma_{\mu 5 S} \otimes \xi_{\mu 5 F}} )_{AB}
\chi(B)\big\}\,.
\label{eq:app:D}
\end{eqnarray}
This also transforms as part of an irrep 
of the timeslice group
[a different irrep from that in which the operator 
(eq.~\ref{eq:truerep}) transforms, unless $\mu=4$].
Thus we can express the HPC operator as a sum of
two true representations:
\begin{eqnarray}
\sum_{\vec n} O^{\rm Klu}_{S,F} &=&
[\sum_{\vec n} \bar{\chi}
( \overline{\gamma_S \otimes \xi_F} )
\chi ]_T
\nonumber \\ & & +
\frac{1}{2} \sum_{\vec n}\bar{\chi}
( \overline{\gamma_{\mu 5 S} \otimes \xi_{\mu 5 F}} )
\overleftrightarrow{D}_\mu \chi 
\end{eqnarray}
Note that for $\mu=4$ these two operators, while having different
spin-taste, lie in the same irrep of the timeslice group.
This is the standard ``doubling'' of states discussed in the text.

The extension to operators containing more than one link is 
straightforward.
For example, for distance two operators ({\em i.e.}
$|S_\mu - F_\mu | = |S_\nu - F_\nu | = 1$ for $\mu \ne \nu$, and
$|S_\rho - F_\rho | = 0$ for $\rho \ne \mu \ne \nu$),
we find:
\begin{eqnarray}
\sum_{\vec n}O^{\rm Klu}_{S,F} &=&
[\sum_{\vec n} \bar{\chi}
( \overline{\gamma_S \otimes \xi_F} )
\chi ]_T
\nonumber \\ & & +
\frac{1}{2} \sum_{\vec n}\bar{\chi}
( \overline{\gamma_{\mu 5 S} \otimes \xi_{\mu 5 F} } )
\overleftrightarrow{D}_\mu \Phi_\nu \chi
\nonumber \\ & & +
\frac{1}{2} \sum_{\vec n}\bar{\chi}
( \overline{\gamma_{\nu 5 S} \otimes \xi_{\nu 5 F} } )
\overleftrightarrow{D}_\nu \Phi_\mu \chi
\nonumber \\ & & +
\frac{1}{4} \sum_{\vec n}\bar{\chi}
( \overline{\gamma_{\mu \nu S} \otimes \xi_{\mu \nu F}} )
\overleftrightarrow{D}_\mu \overleftrightarrow{D}_\nu \chi 
\,.
\end{eqnarray}
Thus there are four irreps (all different if $\mu\ne4\ne\nu$)
contained in the HPC operator.

In general, the HPC operator includes the Golterman
operator (or, if $|S_4-F_4|=1$, an operator in the same irrep)
as the leading term as well as operators with derivatives
having different spin-taste. One expects the latter
to have a coupling to physical states suppressed by
$\sim a\Lambda_{\rm QCD}$ for each derivative.
A similar analysis holds for the cubic U(1) sources,
while the cubic wall sources should produce only true irreps aside
from the effects of using only a few noise vectors.
One should also keep in mind that the symmetries are not
completely implemented when one has a finite ensemble of
gauge configurations, although we expect this effect to be
smaller than the use of only one source-set per configuration.
Thus we expect the contamination from coupling to states of
unwanted spin-taste to be smallest for the combination ``Golterman-wall'', followed by
``Golterman-U(1)'', then ``HPC-wall'', and finally ``HPC-U(1)''. 
This is not, however, the only source of systematic problems.
The coupling to excited states within the same irrep should be more
significant for wall than for U(1) sources.


\end{document}